%% file: main.tex
\documentclass[a4paper]{jpconf}
\usepackage{amssymb}
\usepackage{graphicx}
\usepackage{aas_macros}

\newcommand\unit[2]{\mbox{#1\,#2}}
\newcommand\mathunit[2]{#1\,#2}

\begin{document}

\title{Searches for inspiral gravitational waves associated with short gamma-ray bursts in LIGO's fifth and Virgo's first science run}

\author{ALEXANDER DIETZ on behalf of the LIGO Scientific Collaboration and the Virgo
Collaboration}

\address{LAPP, 9 Chemin de Bellevue,\\
BP 110, 74941 Annecy le Vieux CEDEX, France\\
E-mail: alexander.dietz@lapp.in2p3.fr}

\vspace*{5mm}
\begin{abstract}
Mergers of two compact objects, like two neutron stars or a neutron star and a black hole, are the probable progenitor of short gamma-ray bursts. 
These events are also promising sources of gravitational waves, that are currently motivating related searches by an international network of gravitational wave detectors. 
Here we describe a search for gravitational waves from the in-spiral phase of two coalescing compact objects,
in coincidence with short GRBs occurred during during LIGO's fifth science run and Virgo's first science run.
The search includes 22 GRBs for which data from more than one of the detectors in the LIGO/Virgo network were available. 
No statistically significant gravitational-wave candidate has been found, and a parametric test shows no excess of weak gravitational-wave signals in our sample of GRBs.
The  90\%~C.L. median exclusion distance for GRBs in our sample is of 6.7 Mpc, under the hypothesis of a  neutron star - black hole progenitor model.
\end{abstract}

\section{Introduction}

\vspace*{5mm}
Gamma Ray Bursts (GRB) are intense flashes of $\gamma$-rays traditionally divided in two classes; long duration bursts (duration $\gtrsim$ \unit{2}{s}) which are generally associated with hypernova explosions in star-forming galaxies (see e.g. \cite{RevModPhys.76.1143,meszaros:2006} and references therein) and short duration bursts (duration $\lesssim$ \unit{2}{s}), thought to originate primarily from the coalescence of a neutron star (NS) with another
compact object (see e.g. \cite{NakarReview:2007} and references therein), either a neutron star or a black hole (BH).
Coalescing compact objects are believed to be promising sources of gravitational waves (GW), currently motivating searches by a worldwide network of GW detectors. 
Because the time and sky location of a GRB is known, the search for coincident GW signals can be performed with a reduced threshold, increasing the sensitivity compared to untriggered searches. 
In the analysis presented here, we adopt the GRB classification scheme recently proposed by \cite{0004-637X-703-2-1696}, with a type-I GRB denoting the class of GRBs probably created by a merger (i.e. the 'short' ones), while a type-II GRB denotes the class of GRBs usually associated with stellar core-collapse (a 'long' GRB).

This work describes a search for GW signals associated with type-I GRBs during LIGO's fifth science run (S5) and Virgo's first science run (VSR1). 
S5 took place from 4 November 2005 to 30 September 2007 and VSR1 from 18 May 2007 to 30 September 2007.
S5 comprised of the three LIGO detectors \cite{Abbott:2007kv} which are located at Hanford, WA and Livingston, LA. 
The Hanford site consists of two detectors, one with 4~km long interferometer arms (H1) and one having 2~km long arms (H2), while the detector at Livingston has arms with a length of 4~km (L1). 
VSR1 is comprised of the Virgo detector (V1, \cite{Acernese:2008}), located near Pisa, Italy, with 3~km long arms.

A detection of GWs associated with a type-I GRB would provide direct evidence of the merger nature of such events, having large impact on GRB physics. 
Besides it will be possible to measure component
masses \cite{Cutler:1994,FinnChernoff:1993}, measure component
spins \cite{Poisson:1995ef}, constrain NS equations of
state \cite{flanagan:021502,Read:2009},
test general relativity in the strong-field regime \cite{Will:2005va}, the graviton mass \cite{PhysRevD.80.044002,Keppel:2010qu}  and the  Lorentz Invariance principle \cite{Ellis2006402}. 
Simultaneous observation of a binary's GW signal and type-I GRB will also enable an independent measurement of the binary luminosity distance and redshift, probing the relatively nearby universe's expansion, and complementing other cosmological studies (see e.g. \cite{Nissanke:2009kt} and references therein).

\section{Sample and data selection}
\vspace*{5mm}

The coalescence model suggests that the time delay between the arrival of the GW and the subsequent electromagnetic burst, referred to as trigger time, is of order one second or less. 
If the initial spike of GRBs is created by so-called internal shocks (produced by the impact of faster outmoving shells on slower moving shells in the 'standard model' of GRBs (see e.g. \cite{meszaros:2006} and references therein), the time-delay is on the millisecond scale \cite{meszaros:2006,NakarReview:2007} or at least below 1 second \cite{Finn:2004}. 
With a more semi-analytical description of the  final stage of a NS-–BH merger it has been calculated that the majority of matter plunges on the BH within 1 second \cite{davies:2005} and numerical simulations on the mass transfer suggest a timescales of milliseconds \cite{rosswog-2005} or some seconds at maximum \cite{Faber:2006tx}. 
These considerations prompted the use of a \unit{6}{s} long time window (\textit{on-source}) around the trigger time (from -5 to +1), in which we expect the GW signal to be.

Data around this on-source segment is chosen to estimate the background and the efficiency of the search, split up into 324 segments\footnotemark\footnotetext{corresponding to the minimum amount of data required for a standard analysis} with a length of \unit{6}{s} each, not necessarily symmetrical around the on-source segment. To prevent a possible loud on-source signal biasing the background estimation, we discard \unit{48}{s} on each side of the on-source segment, which reflects the length of the longest used template for this search; the remaining segments constitute the \textit{off-source} segments.
Considering a padding time of \unit{72}{s} on both sides of the data stretch, the total amount of data used for the analysis is \unit{2190}{s} for each GRB.
To ensure the use of only good quality data, any \unit{6}{s} long segment which overlaps a data-quality veto is discarded \cite{abbott-2009}. 
Therefore the actual used number of background trials might differ from 324.  

\begin{table}
\begin{center}
\begin{tabular}{lccl}
\hline\hline
 GRB            & Redshift  & Duration (s)      & References\\\hline
 051114 & \ldots          &  2.2 & \cite{gcn4272,gcn4275} \\
 051210 & \ldots          &  1.2 & \cite{gcn4315,gcn4321} \\
 051211 & \ldots          &  4.8 & \cite{gcn4324,gcn4359} \\
 060121 & \ldots          &  2.0 & \cite{gcn4550} \\
 060313 & $<$1.7 &  0.7 & \cite{gcn4867,gcn4873,gcn4877} \\
 060427B & \ldots          &  2.0 & \cite{gcn5030} \\
 060429 & \ldots          &  0.25 & \cite{gcn5039} \\
 061006 & \ldots          &  0.50 & \cite{gcn5699,gcn5704} \\
 061201 & \ldots          &  0.80 & \cite{gcn5881,gcn5882} \\
 061217 & 0.827 &  0.30 & \cite{gcn5926,gcn5930,gcn5965} \\
 070201 & \ldots          &  0.15 & \cite{gcn6088,gcn6103} \\
 070209 & \ldots          &  0.10 & \cite{gcn6086} \\
 070429B & \ldots          &  0.50 & \cite{gcn6358,gcn6365} \\
 070512 & \ldots          &  2.0 & \cite{gcn6408} \\
 070707 & \ldots          &  1.1 & \cite{gcn6605,gcn6607} \\
 070714 & \ldots          &  2.0 & \cite{gcn6622} \\
 070714B & 0.92 &  64.0 & \cite{gcn6620,gcn6623,gcn6836} \\
 070724 & 0.46 & 0.40 & \cite{gcn6654,gcn6656,gcn6665} \\
 070729 & \ldots          &  0.90 & \cite{gcn6678,gcn6681} \\
 070809 & \ldots          & 1.3 & \cite{gcn6728,gcn6732} \\
 070810B & \ldots          &  0.08 & \cite{gcn6742,gcn6753} \\
 070923 & \ldots          &  0.05 & \cite{gcn6818,gcn6821} \\ \hline
\end{tabular}
\end{center}
\caption{Parameters of the 22 GRBs selected for this search. Two of the GRBs in this list have a $T_{90}$ duration larger than \unit{2}{s}, but spectral features suggest a type-I GRB affiliation. }
\label{tab:listGRB}
\end{table}

During the whole period of S5 and VSR1, 212 GRBs in total were observed, mainly with the \emph{SWIFT} and HETE missions, but also with the Interplanetary Network (IPN) \cite{2003AIPC..662....3Rx,Gehrels:2004,1995SSRv...71..265A,2003A&A...411L...1W,Atwood:2009ez}.
33 of them has been found to belong to the type-I GRB class.
Because we require coincident data from at least two detectors, only 22 GRBs have been selected for the analysis (see Table~\ref{tab:listGRB}).
If data from three or more detectors were available, only data from the two most sensitive detectors has been used for the analysis, except for GRB~070923 (see below). 
  Two of these 22 selected GRBs actually had a $T_{90}$ duration larger than 2~s, but spectral features suggest their affiliation to the type-I GRB class. 
There are 9 GRBs in H1--H2 coincident time, 11 in H1--L1, 1 in H2--L1, and 1 GRB in H1--L1--V1 coincident time. 
The latter, GRB~070923, is a special case because it uses data from three spatially separated detectors, which all have about the same directional sensitivity\footnotemark\footnotetext{taking into account the response of the detector in the direction of the GRB} to this GRB.
Worth special mention is also GRB~070201, which has been already analysed in a high-priority search because the progenitor was possibly located in M31, a galaxy only $\sim$780~kpc away. 
No GW signal was found and a coalescence scenario was ruled out with $>$99\% confidence at that distance \cite{GRB070201}, hinting to a soft-gamma repeater progenitor scenario for this type-I GRB in M31 \cite{0004-637X-680-1-545, 0004-637X-681-2-1464}.
Nevertheless, this GRB is being reanalysed in the current work because of the use of a lower threshold and the use of a more sensitive likelihood-ratio based ranking method. 

\section{Search details}
\vspace*{5mm}

\begin{table*}
\begin{center}
\begin{tabular}{l|cccc|c|ccc}
\hline\hline
& \multicolumn{4}{c|}{Antenna response} & & \multicolumn{2}{c}{Excluded distance (Mpc)} \\
GRB & H1 & H2 & L1 & V1 & F.A.P. & NS--NS & NS--BH \\ \hline
\input{table} \hline
\end{tabular}
\end{center}
\caption{Summary of the results for the search for GWs
from each GRB. The Antenna Response column contains
the response for each detector in the direction of the source; a value of 1 corresponds to optimal sensitivity and a value of 0 corresponds to no sensitivity in that direction (see \cite{S5GRB} for details). An ellipsis (\ldots) denotes that a detector's
data were not used. F.A.P. is the false-alarm
probability of the most significant on-source candidate; trials with no candidates above threshold are assigned a F.A.P. of 1.
The last two columns show the 90\% CL exclusion distances to the assumed merger source. }
\label{tab:listResults}
\end{table*}

The inspiral waveforms generated by coalescing NS-NS or BH-NS systems, can be reliably predicted using post-Newtonian perturbation theory, until the last fraction of a second prior to merger. These waveforms can be used in matched filtering of noisy data from GW detectors to identify GW candidate events. In this search, data are analysed using the standard compact binary search pipeline described in detail in  \cite{Collaboration:2009tt}, which is indeed based on a matched-filter technique to filter the data against post-Newtonian approximants of the expected GW signal.
First a discrete bank of template waveforms is set up that span a
two-dimensional parameter space (one for each component mass) such
that the maximum loss in signal to noise ratio (SNR) for a binary with
negligible spins would be $3\%$ \cite{hexabank}.
While this template bank does not contain waveforms with spin, we verified that the search can still detect binaries with reasonable spin orientations and magnitudes. 
The template bank spans the component masses symmetrically in the range  $[\mathunit{1}, \mathunit{40}) {M_\odot}$ in total mass. 

Then the data from each of the used detectors is filtered with each of the template, and a \textit{trigger} is generated when the matched filter SNR exceeds a specific threshold. 
The thresholds were selected specifically for each detector combination to reflect the individual directional sensitivity of a detector;
details can be found in \cite{S5GRB}. 
We also applied two signal-based tests to reduce the number of background triggers. 
The first test ($\chi^2$ test \cite{Allen:2004}) rejects triggers with too large differences between the expected and measured amplitude series of the signal, while the second test ($r^2$ test \cite{Rodriguez:2007}) discards triggers depending on how long the $\chi^2$ value stays above a certain threshold around the trigger time. 
Finally, coincident triggers, found to be consistent in their time and mass parameters from two or more detectors \cite{Robinson:2008}, are recorded as \textit{candidate events}.

\section{Likelihood ratio ranking}
\label{sec:likelihoof}
\vspace*{5mm}

To obtain the search efficiency and to calculate the likelihood ratio of a candidate event, simulated waveform signals (\textit{software injections}) are added to the off-source data and processed with the pipeline described in the previous section.
These so-called \textit{software injections} are made over a wide range of parameters (i.e. masses and spins of the two objects, and inclination angle and polarization angle of the orbital axis, describing the orientation of the binary with respect to the line-of-sight to earth) to assess the full efficiency.
The number of recovered injections compared to the total number of injections determined the efficiency of the search, as a function of the various parameters. 

Another quantity determined by the search is the false-alarm probability; this is the probability that a given candidate is found by pure chance and is obtained by the relative number of off-source trials yielding a statistical significance equal or greater than the on-source measurement.
With these two quantities we define a \textit{likelihood ratio} of a candidate, which is simply the efficiency divided by the false-alarm probability. The larger this likelihood ratio, the more likely it is that this candidate is a real GW.

\begin{figure}[tbp]
\begin{center}
\includegraphics{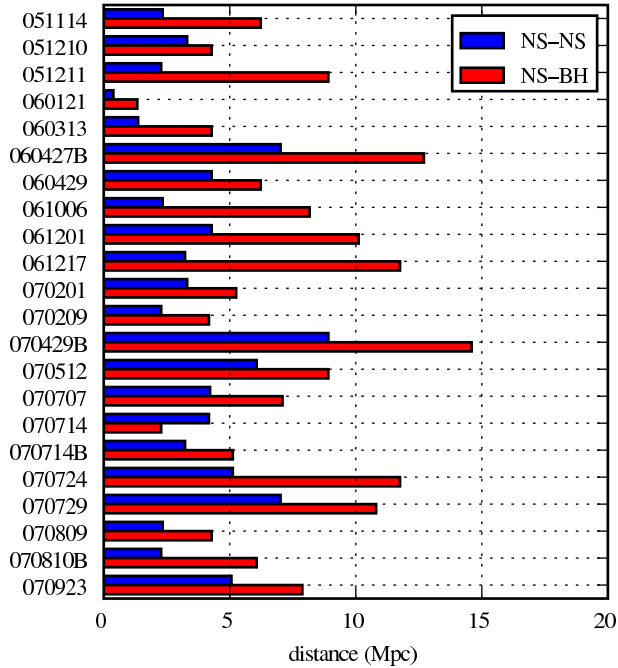}
\end{center}
\caption{\label{fig:exclusion}The 90\% C.L. exclusion distances to putative NS-NS and NS-BH progenitor systems for each analysed SGRB as listed in Table \ref{tab:listGRB}. The median exclusion distance for a  NS-NS is \unit{3.3}{Mpc} and is \unit{6.7}{Mpc} for a NS-BH system.}
\end{figure}

The analysis yielded no evidence for a GW signal in coincidence
with a GRB in our sample; see Table \ref{tab:listResults} for detailed results. With the null observations and the large number of
injection trials we were able to constrain the distance to each GRB assuming it was
caused by a compact binary coalescence.

We used the approach of Feldman-Cousins \cite{Feldman:1997qc} to compute regions in distance where GW
events would, with a given confidence, have produced results inconsistent with our observations, for a given companion mass range. 
We quote the median exclusion distances for two mass ranges, representing the merger of two neutron stars (NS-NS), with a component mass in the range $\mathunit{[1,\,4)}{M_\odot}$, and for a merger of a neutron star and a black hole (NS-BH), with the black hole mass in the range $\mathunit{[7,\,10)}{M_\odot}$.
The median exclusion distance for the NS-NS systems is \unit{3.3}{Mpc} while for the more massive NS-BH systems the median exclusion distance is found to be \unit{6.7}{Mpc};  
more details can be found in~\cite{S5GRB}.
A graphical representation of the individual excluded distances can be seen in Fig. \ref{fig:exclusion}.

The most significant candidate was GRB~070201 with a false-alarm probability of 6.8~\%.
This candidate was found to be just below the threshold used in the previous analysis  \cite{GRB070201}, which showed no candidate in the on-source segment at all.
Despite the low  false-alarm probability, this candidate is consistent with the expectation from the off-source segments.

\section{Population test}
\vspace*{5mm}
\label{sec:pop}

In addition to the individual detection searches, we assess the presence of a GW signal too weak to stand out above background separately, but which is significant when the entire sample of analysed GRBs are taken together.
This is being done with the non-parametric Wilcoxon-Mann-Whitney U test \cite{MannWhitney}, to answer the question of whether both samples (significances of the on-source and off-source segments) are drawn from the same population. 
This test calculates a U statistic by the sums of the ranks of both samples, and compares it to the expected value if both samples were drawn from the same distribution. 
Since this statistic is approximately Gaussian distributed, a one-sided probability $p$ can be calculated that either shows the two samples are indeed drawn from the same distribution ($p\simeq50\%$) or contain a population of weak signals in the sample ($p\lesssim5\%$). 
A large value for $p$ would hint at a problem with the analysis, since it implies more significant candidates in the off-source than in the on-source.

Applying the U test, we find that the two distributions are consistent with each other; if the
on-source and off-source significances were drawn from the same distribution, they would yield a U statistic greater than what we observed 53~\% of the time.
Therefore we find no evidence for an excess of weak GW signals associated with GRBs in our sample.

\section{Summary and future work}
\vspace*{5mm}

We presented a search for a merger signal in data around 22 type-I GRBs, probably created by a merger of two compact objects. 
This analysis was based on the standard matched filtering techniques used also for untriggered searches, but with a lower threshold. 
The analysis also used a more sensitive likelihood ratio based ranking technique to determine the significance of the candidate events. 
No GW signal in data around any type-I GRB in our sample has been found, and exclusion distances to each of the GRB has been computed. The median exclusion distance is  \unit{3.3}{Mpc} for a NS-NS system, and \unit{6.7}{Mpc} for a NS-BH system.
A non-parametric Wilcoxon-Mann-Whitney U test has been applied to the sample of on-source candidates, which was found to be consistent with the off-source background candidates. 

With the conclusion of LIGO's fifth and Virgo's first science runs, the detectors have undergone a series of enhancements, followed by new data
taking periods, started in summer 2009 with S6 and VSR2 runs, alternating with interruptions for further enhancements and commissioning activity.
A fully automated online search for merger signal from type-I GRBs has been set up for this period, analysing the data around each trigger in the same way as it has been described in this paper. 
Regarding the analysis, several short-term improvements are scheduled, such as an improvement of the background estimation, a faster analysis of the data and the implementation of a coherent search stage; the results of this analysis will be published elsewhere.

These data taking periods will last until end 2010 - mid 2011, when a shutdown of the detectors is planned, in order to implement advanced detector configurations \cite{1742-6596-32-1-033,0264-9381-27-8-084006}.
With these detectors, which will start to be operated in $\sim$2014-2015, the sensitivity will gradually improve, aiming at a factor of $\sim$~10 improvement with respect to the initial configuration, increasing the observable volume of the universe by a factor of $\sim$~1000. 
By then, coincident detections of GWs with type-I GRBs are expected to be very likely \cite{ratesdoc}, marking the beginning of exciting gravitational-wave physics.

\section*{Acknowledgements}

The authors gratefully acknowledge the support of the United States
National Science Foundation for the construction and operation of the
LIGO Laboratory, the Science and Technology Facilities Council of the
United Kingdom, the Max-Planck-Society, and the State of
Niedersachsen/Germany for support of the construction and operation of
the GEO600 detector, and the Italian Istituto Nazionale di Fisica
Nucleare and the French Centre National de la Recherche Scientifique
for the construction and operation of the Virgo detector. The authors
also gratefully acknowledge the support of the research by these
agencies and by the Australian Research Council, the Council of
Scientific and Industrial Research of India, the Istituto Nazionale di
Fisica Nucleare of Italy, the Spanish Ministerio de Educaci\'on y
Ciencia, the Conselleria d'Economia Hisenda i Innovaci\'o of the
Govern de les Illes Balears, the Foundation for Fundamental Research
on Matter supported by the Netherlands Organisation for Scientific Research,
the Polish Ministry of Science and Higher Education, the FOCUS
Programme of Foundation for Polish Science,
the Royal Society, the Scottish Funding Council, the
Scottish Universities Physics Alliance, The National Aeronautics and
Space Administration, the Carnegie Trust, the Leverhulme Trust, the
David and Lucile Packard Foundation, the Research Corporation, and
the Alfred P. Sloan Foundation.
This document has been assigned LIGO document number P1000046.

\vspace*{1cm}
\bibliographystyle{iopart-num}
\input{main.bbl}

\end{document}

%% file: table.tex
GRB~051114 &0.56 & 0.56 & \ldots & \ldots & 1 & 2.3 & 6.2 \\
GRB~051210 &0.61 & 0.61 & \ldots & \ldots & 0.10 & 3.3 & 4.3 \\
GRB~051211 &0.53 & \ldots & 0.62 & \ldots & 0.66 & 2.3 & 8.9 \\
GRB~060121 &0.11 & \ldots & 0.09 & \ldots & 0.58 & 0.4 & 1.3 \\
GRB~060313 &0.59 & 0.59 & \ldots & \ldots & 0.16 & 1.4 & 4.3 \\
GRB~060427B &0.91 & \ldots & 0.92 & \ldots & 1 & 7.0 & 12.7 \\
GRB~060429 &0.92 & 0.92 & \ldots & \ldots & 0.21 & 4.3 & 6.2 \\
GRB~061006 &0.61 & 0.61 & \ldots & \ldots & 1 & 2.3 & 8.2 \\
GRB~061201 &0.85 & 0.85 & \ldots & \ldots & 1 & 4.3 & 10.1 \\
GRB~061217 &0.77 & \ldots & 0.52 & \ldots & 0.23 & 3.2 & 11.8 \\
GRB~070201 &0.43 & 0.43 & \ldots & \ldots & 0.07 & 3.3 & 5.3 \\
GRB~070209 &0.19 & \ldots & 0.12 & \ldots & 0.76 & 2.3 & 4.2 \\
GRB~070429B &0.99 & \ldots & 0.93 & \ldots & 0.31 & 8.9 & 14.6 \\
GRB~070512 &0.38 & \ldots & 0.51 & \ldots & 0.97 & 6.1 & 8.9 \\
GRB~070707 &\ldots & 0.87 & 0.79 & \ldots & 0.87 & 4.2 & 7.1 \\
GRB~070714 &0.28 & \ldots & 0.40 & \ldots & 0.72 & 4.2 & 2.3 \\
GRB~070714B &0.25 & \ldots & 0.38 & \ldots & 0.54 & 3.2 & 5.1 \\
GRB~070724 &0.53 & \ldots & 0.70 & \ldots & 0.84 & 5.1 & 11.8 \\
GRB~070729 &0.85 & 0.85 & \ldots & \ldots & 0.40 & 7.0 & 10.8 \\
GRB~070809 &0.30 & 0.30 & \ldots & \ldots & 1 & 2.3 & 4.3 \\
GRB~070810B &0.55 & \ldots & 0.34 & \ldots & 0.50 & 2.3 & 6.1 \\
GRB~070923 &0.32 & \ldots & 0.40 & 0.69 & 0.74 & 5.1 & 7.9 \\

%% file: main.bbl
\providecommand{\newblock}{}